\documentclass{emulateapj}

\newcommand{\mbh}{$M_{\rm BH}$}
\newcommand{\sigs}{$\sigma_{\ast}$}
\newcommand{\msig}{\mbh-\sigs}
\newcommand{\sigo}{$\sigma_{\rm [O\, III]}$}

\shorttitle{BH MASSES OF SOFT X-RAY-SELECTED AGNs}

\begin{document}

\title{Revisiting the Black Hole Masses of Soft X-ray-Selected Active
  Galactic Nuclei} 

\author{Linda C. Watson \altaffilmark{1}, Smita Mathur \altaffilmark{1},
  Dirk Grupe \altaffilmark{2}
}

\email{watson@astronomy.ohio-state.edu}

\altaffiltext{1}{Department of Astronomy, The Ohio State University,
  Columbus, OH, 43210, USA}
\altaffiltext{2}{Department of Astronomy and Astrophysics,
  Pennsylvania State University, University Park, PA 16802} 

\begin{abstract}
In our previous work, using luminosity and the H$\beta$ FWHM as
surrogates for black hole mass ($M_{\rm BH}$), we compared the black
hole masses of narrow-line Seyfert 1 galaxies (NLS1s) and broad-line
Seyfert 1 galaxies (BLS1s) in a sample of soft X-ray-selected active
galactic nuclei. We found that the distributions of black hole masses
in the two populations are statistically different.  Recent work shows
that the second moment of the H$\beta$ emission line (the line
dispersion) is a better estimator of black hole mass than FWHM. To
test whether changing the width measure affects our results, we
calculate line dispersion-based black hole masses for our soft X-ray
selected sample. We find that using the line dispersion rather than
the FWHM as a measure of the gas velocity shifts NLS1 and BLS1 virial
product distributions closer together, but they remain distinct. On
the \msig\ plane, we find that using the line dispersion leaves NLS1s
below the \msig\ relation, but to a less significant degree than when
FWHM is used to calculate black hole masses (the
[\ion{O}{3}]$\,\lambda$5007 FWHM is used as a surrogate for the bulge 
stellar velocity dispersion). The level of significance of our
findings is such that we cannot draw firm conclusions on the location
of the two samples on the \msig\ plane. We are still left with two
alternative scenarios: either (1) NLS1s lie below the \msig\ relation
indicating that their black hole masses are growing, or (2) NLS1s lie
on the \msig\ relation, so they preferentially reside in smaller mass,
less luminous galaxies; the present data do not allow us to choose one
over the other. More trustworthy stellar velocity dispersions and
accurate black hole mass measurements with reverberation mapping are
required for a firmer statement about the locus of NLS1s on the \msig\
plane. 
\end{abstract}

\keywords{galaxies: active --- galaxies: nuclei --- quasars: general}

\section{Introduction}

In the study of the co-evolution of host galaxies and their central
black holes, one of the most interesting and useful tools is the firm
correlation between black hole mass ($M_{\rm BH}$) and stellar velocity
dispersion of the host bulge ($\sigma_{\ast}$).  The \msig\ relation was
first established for quiescent galaxies, where black hole masses were
mainly determined using gas and stellar dynamics \citep{ferrarese1,
gebhardt1} and in few cases with maser kinematics (e.g. NGC~4258;
Miyoshi et al. 1995) and proper motion (Galactic center: Genzel et
al. 2000, Ghez et al. 2000).  With further study, the \msig\ relation
was found to extend to active galaxies, where traditional mass
measurement techniques are no longer feasible because the region
directly affected by the black hole is unresolved \citep{gebhardt2,
ferrarese2}.  For type 1 active galactic nuclei (AGNs), the most
direct technique for measuring black hole masses is reverberation
mapping \citep{blandford, brad93}. Here the time delay between
continuum and associated emission line variations is used with the
width of the emission line to calculate a virial mass.  However,
reverberation mapping is time intensive, and thus a common practice
in determining black hole masses is to employ an empirical relation
between the radius of the broad line region ($R_{\rm BLR}$) and the
monochromatic continuum luminosity ($L_{5100}$), as derived in, e.g.,
\citet{kaspi} or \citet{misty}.  The $R_{\rm BLR}-L_{5100}$ relation
is calibrated against reverberation mapped AGNs and allows one to
calculate $R_{\rm BLR}$ and therefore the virial mass using a
single-epoch observation.  Consequently, black hole masses can be
estimated for large samples of AGNs.

This contribution continues the three-paper series investigating the
locus of narrow-line Seyfert 1 galaxies (NLS1s) on the \msig\ plane
\citep{paper1, paper2, paper3}.  NLS1s are defined as those AGNs
having an H$\beta$ emission line ${\rm FWHM} \leq 2000\, {\rm km\ s}^{-1}$
\citep{pogge}.  The comparatively low emission line widths of NLS1s
are commonly accepted as evidence for a low mass black hole powering
the AGN and often go hand-in-hand with a high accretion rate and a
steep soft X-ray slope \citep{pounds, grupe98}. NLS1s are therefore a
focal point for extreme AGN physical properties and also occupy a
unique and interesting position on the \msig\ relation.

\citet{paper1} studied a sample of 75 soft X-ray-selected Seyfert 1
galaxies, including 32 NLS1s and 43 broad-line Seyfert 1 galaxies
(BLS1s), extending the earlier work of \citet{mathur01}.  Using the
H$\beta$ FWHM and the \citet{kaspi} $R_{\rm BLR}-L_{5100}$ relation to
estimate the black hole mass and the width of the
[\ion{O}{3}]$\,\lambda$5007 emission line ($\sigma_{\rm [O\, III]}$)
as a surrogate for the stellar velocity dispersion, they found that
NLS1s as a class lie below the \msig\ relation.  Furthering the study,
\citet{paper2} distinguished between those NLS1s well below the \msig\
relation and those near the \msig\ relation and found that those NLS1s
that lie below the \msig\ relation also have larger Eddington ratios
($L_{\rm bol}/L_{\rm Edd}$) and steeper soft X-ray slopes compared to
those NLS1s that lie near the \msig\ relation.  Both results led the
authors to conclude that highly accreting AGNs at low redshift lie
below the \msig\ relation while AGNs with low accretion rates lie
close to the \msig\ relation, having achieved their final black hole
mass.

In \citet{paper1}, neither the black hole masses based on H$\beta$
FWHM nor the stellar velocity dispersions based on \sigo\ are direct
measurements.  Therefore, the most prudent approach is to test that
both H$\beta$ FWHM and \sigo\ accurately describe \mbh\ and
$\sigma_{\ast}$, respectively.  By far the most suspect of the two
estimations is substituting \sigo\ for \sigs\ \citep{boroson,
greene05}.  Consequently, \citet{paper3} addresses this point by
focusing on concerns put forth by \citet{greene05} that \sigo\ is not
only a function of $\sigma_{\ast}$, but also of $L_{\rm bol}/L_{\rm
Edd}$.  Correcting for this dependence does not change the results of
\citet{paper1}: while no individual object's location on their $M_{\rm
BH}$-\sigo\ plot should necessarily be trusted, the overall result
is sound in that highly accreting NLS1s lie below the \msig\
relation \citep{paper3}. 

In this work, we address the \mbh\ estimates for the soft X-ray 
selected sample of \citet{paper1}.  The motivations for this study are the 
recent publications by \citet{brad04} and \citet{brad}.  Our discussion is 
based on black hole masses ($M_{\rm BH}$) calculated using
\begin{equation}
\label{eqn:virial}
M_{\rm BH}=f \frac{R_{\rm BLR} (\Delta V)^{2}}{G},
\end{equation}
where $f$ is a scale factor that depends on the geometry and
kinematics of the broad line region (BLR), $R_{\rm BLR}$ is the radius
of the BLR, and $\Delta V$ is a measure of the BLR gas velocity.  We
will often refer to the virial product [${\rm VP}=R_{\rm BLR}(\Delta
V)^{2}/G$], which only differs from \mbh\ by the dimensionless scale
factor $f$, which is expected to be of order unity.  \citet{brad04}
measured the emission line widths and the time delays between
continuum and line variations ($\tau$) for various emission lines in
four reverberation mapped AGNs. They found that using the second
moment of the H$\beta$ emission line, referred to as the ``line
dispersion'' or $\sigma_{\rm line}$, as a measure of $\Delta V$
reproduces a $\Delta V\propto\tau^{-1/2}$ relation with higher
precision than FWHM. They therefore conclude that the line dispersion
is a more robust width measure than FWHM because it provides a more
constant virial product over multiple observations of an AGN.

In \citet{brad}, the authors divided their sample of 14 AGNs in two  
ways: first with Population 1 having H$\beta$  FWHM/$\sigma_{\rm line}$ 
$<$ 2.35 and Population 2 having FWHM/$\sigma_{\rm line}$ $>$ 2.35, and 
second with Population A having H$\beta$ FWHM $<$ 4000 ${\rm km\ s}^{-1}$  
and Population B having FWHM $>$ 4000 ${\rm km\ s}^{-1}$. Generally,
Population 1  and A are considered narrow-line objects and Population
2 and B are considered broad-line objects. They then calculated virial
products using both FWHM and $\sigma_{\rm line}$ as $\Delta V$.  To
determine the statistical value of the scale factor $f$ for each of
the four populations above, they shifted these virial products onto
the quiescent galaxy \msig\ relation of \citet{tremaine} \citep[this
is the method of][]{onken}.  When using the H$\beta$ FWHM as $\Delta
V$, the scale factors they derive for narrow-line objects and
broad-line objects are significantly different.  On the other hand,
the scale factors computed using the H$\beta$ line dispersion are
consistent with a constant value.  \citet{brad} therefore conclude
that the line dispersion is less sensitive to whatever property it is
that establishes a difference between narrow and broad populations in  
the eyes of FWHM, and is therefore a less biased width measure.  The
authors provide their best estimates for these various scale factors;
the one of most interest for this work is $f=3.85$, which was derived
to convert line dispersion-based virial products measured on the mean
spectrum into black hole masses.

We will begin this work by comparing NLS1 and BLS1 virial product
distributions.  When we require masses, we will adhere to the approach
of \citet{paper1} by examining the positions of NLS1s and BLS1s with
respect to the \msig\ relation under the assumption that a single
scale factor is appropriate for both populations.  In
Section~\ref{sec:analysis}, we discuss our data analysis technique,
detailing our line dispersion measurements. Section~\ref{sec:results}   
describes our results, comparing NLS1 and BLS1 virial product and
\msig\ distributions when the H$\beta$ FWHM or line dispersion is used
for $\Delta V$.  Finally, we discuss our conclusions in
Section~\ref{sec:discussion}.

\section{Data Analysis}
\label{sec:analysis}

\citet{paper1} calculated black hole masses for 75 soft X-ray-selected
AGNs (32 NLS1s and 43 BLS1s) using Equation \ref{eqn:virial}.  For $f$
they used the \citet{kaspi} value of 0.75, for $\Delta V$ they used
the H$\beta$ FWHM, and for $R_{\rm BLR}$ they used the \citet{kaspi}
$R_{\rm BLR}-L_{5100}$ relation. Motivated by \citet{brad04} and
\citet{brad}, in this work we use the same sample but calculate virial
products using the H$\beta$ line dispersion rather than FWHM.  We also
update our $R_{\rm BLR}-L_{5100}$ calculation by employing the more recent
\citet{misty} relation.

We carried out measurements on the same H$\beta$ narrow component- and
\ion{Fe}{2}-subtracted spectra as in \citet{grupe} and \citet{paper1}.
We removed four AGNs (one NLS1 and three BLS1s) from our sample
because H$\beta$ was possibly contaminated by an optically thin, very
broad H$\beta$ component \citep{shields}, residual \ion{Fe}{2}, or
\ion{He}{2}.  For consistency with \citet{brad} and to avoid blending
conflict with the [\ion{O}{3}]$\, \lambda\lambda$4959, 5007 lines and 
\ion{Fe}{2}, we chose to measure the line dispersion using the blue
side of the emission line, thus assuming a symmetric line profile.  We
then measured the H$\beta$ line dispersion and FWHM for each AGN.  We
found general agreement between our FWHM measurements and
\citet{paper1} FWHM measurements, signaling that our line dispersion
values can also be trusted. We calculated virial products based on
these line dispersion measurements and compared them to virial
products based on the 2004 FWHM measurements.

Following the procedure of \citet{paper1}, we use the width of the
[\ion{O}{3}]$\,\lambda$5007 emission line as a surrogate for the
stellar velocity dispersion.  We have used twice the half width at
half maximum (HWHM) of the red side of the [\ion{O}{3}] emission line
rather than the FWHM to avoid the blue asymmetry discussed in
\citet{paper1}. Then, $\sigma_{\rm [O\, III]} = 2\cdot {\rm
  HWHM}/2.35$.  We do not present stellar velocity dispersions with
the \citet{greene05} correction to \sigo\ applied because it does not
significantly affect the results (Mathur \& Grupe 2005b). 

\begin{figure*}
\figurenum{1}
\plotone{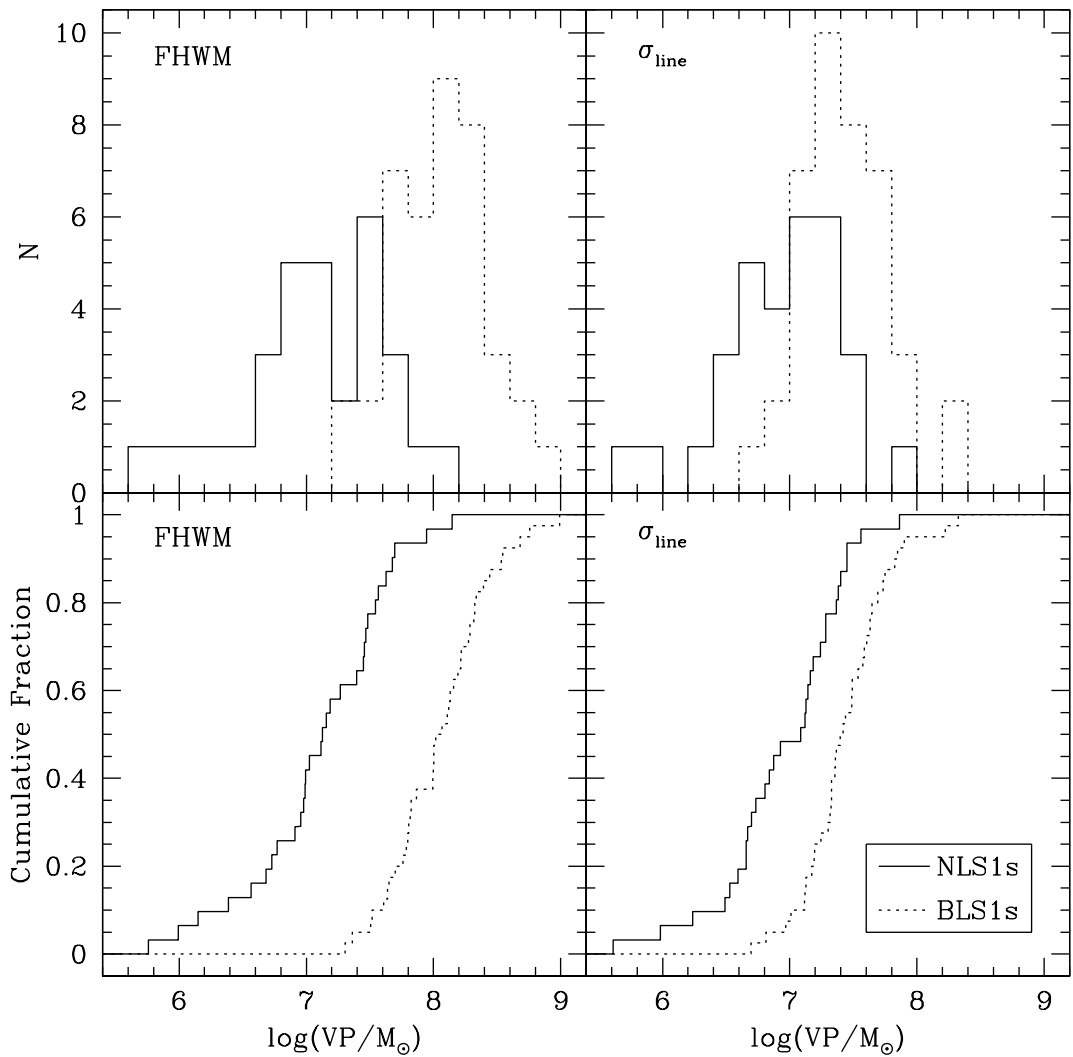}
\label{fig:VP_comp}
\caption{{\it Top panels}: Virial product histograms for NLS1s ({\it
  solid line}) and BLS1s ({\it dotted line}).  {\it Bottom panels}:
  Cumulative fractions of a K-S test comparing NLS1 and BLS1 virial
  product distributions.  In the left panels, we calculate the
  virial product using the H$\beta$ FWHM for $\Delta V$ of
  eq.~\ref{eqn:virial}.  In the right panels, we calculate the
  virial product using the H$\beta$ line dispersion ($\sigma_{\rm
  line}$) for $\Delta V$.  In all panels of this figure and in all
  subsequent figures, we use the \citet{misty} radius-luminosity
  relation to calculate $R_{\rm BLR}$ of eq.~\ref{eqn:virial}. \\
  } 
\end{figure*}

\section{Results}
\label{sec:results}

\subsection{Virial Product Distributions}

In all figures, we have used the \citet{misty} $R_{\rm BLR}-L_{5100}$
relation to calculate $R_{\rm BLR}$ of the virial product.  But for
ease of comparison  with the results of \citet{paper1}, we will quote
results using the \citet{kaspi} $R_{\rm BLR}-L_{5100}$ relation as well.

Also for comparison purposes, we will present figures and
calculations where $\Delta V$ of Equation~\ref{eqn:virial} is the
H$\beta$ FWHM of \citet{paper1} alongside figures where $\Delta V$ is
the H$\beta$ line dispersion.  In the FWHM scheme, the top left panel
of Figure~\ref{fig:VP_comp} shows virial product histograms for NLS1s
({\it solid line}) and BLS1s ({\it dotted line}).  The conclusion that
these distributions are dissimilar is emphasized by the virial product 
cumulative fraction plot in the bottom left panel of
Figure~\ref{fig:VP_comp}.  The Kolmogorov-Smirnov (K-S) test 
probability that our NLS1 and BLS1 FWHM-based virial products are
drawn from the same parent population is $\sim10^{-9}$ (when
the Kaspi et al$.$~2000 $R_{\rm BLR}-L_{5100}$ relation is used, the
probability is $\sim10^{-8}$).

\begin{figure*}
\figurenum{2}
\plotone{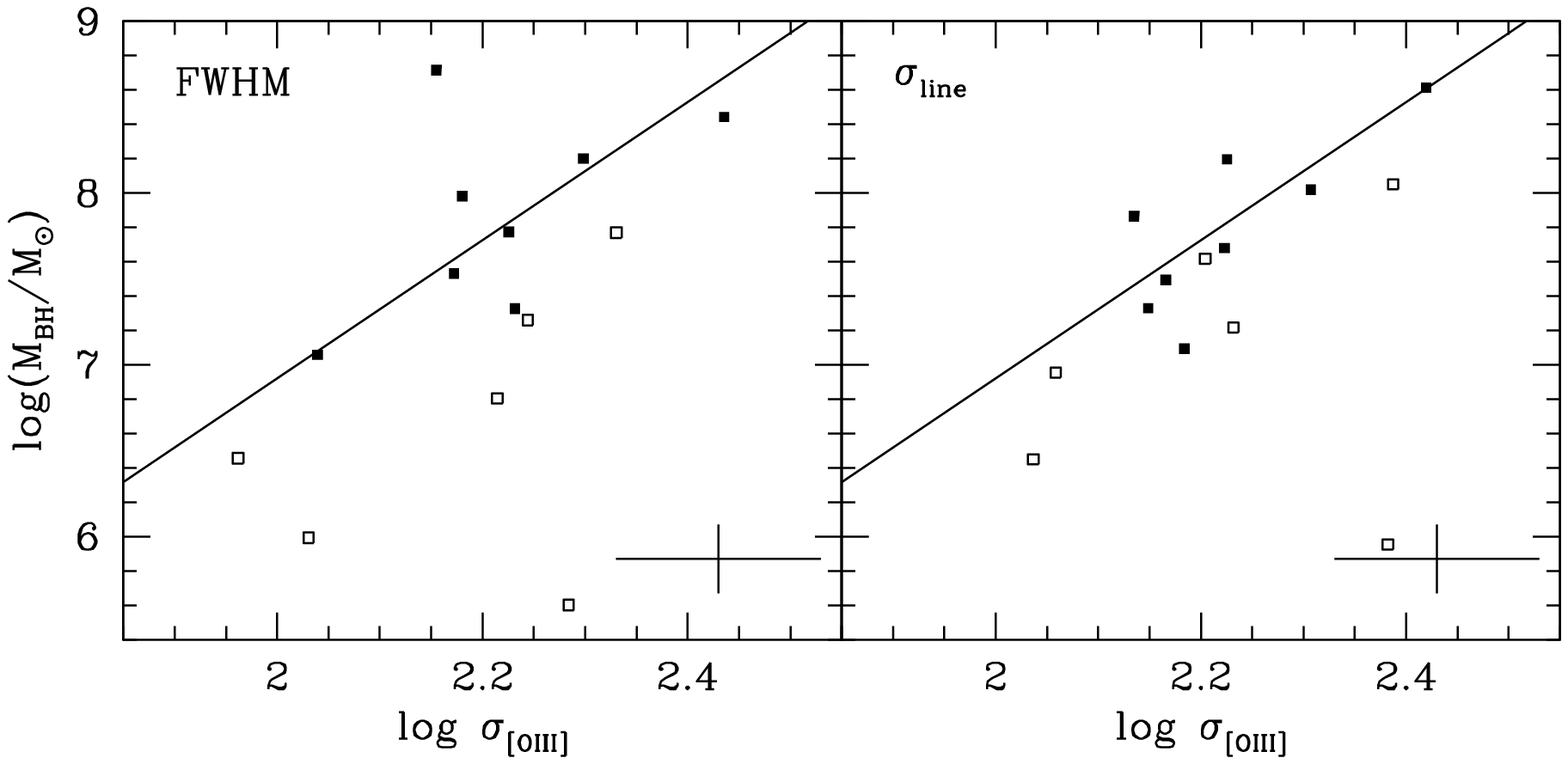}
\label{fig:msig}
\caption{Black hole mass vs. stellar velocity dispersion.  The squares
  show the mean ${\rm log}\, \sigma_{\rm [O\, III]}$ for bins in
  log($M_{\rm BH}/M_{\odot}$), with open squares referring to NLS1s
  and filled squares referring to BLS1s. $\sigma_{\rm [O\, III]}$ has
  units of kilometers per second here.  The solid lines denote the
  relation of \citet{tremaine}.  In the left panel, we calculate
  black hole mass using the H$\beta$ FWHM for $\Delta V$ and a scale
  factor of $f=0.53$. In the right panel, we calculate black hole
  mass using the H$\beta$ line dispersion and a scale factor of
  $f=2.19$.  In each panel, we use the width of the
  [\ion{O}{3}]$\,\lambda$5007 emission line as a surrogate for the
  stellar velocity dispersion.  A typical error bar is shown in the
  lower right corner of each panel, where we have reduced the error in
  the original data in accordance with the binning. \\ }
\end{figure*}

Now using our H$\beta$ line dispersion measurements for $\Delta V$ of
Equation~\ref{eqn:virial}, the top right panel of
Figure~\ref{fig:VP_comp} shows virial product histograms for NLS1s and
BLS1s.  The bottom right panel of Figure~\ref{fig:VP_comp} shows the
virial product cumulative fraction plot.  By comparing the histograms
and cumulative fraction plots of Figure~\ref{fig:VP_comp}, we see that
using the line dispersion closes the gap between NLS1 and BLS1 virial
product distributions.  When the line dispersion is used for $\Delta V$,
the probability that our NLS1 and BLS1 samples are drawn from the same
virial product population is $\sim10^{-4}$ (when the Kaspi et
al$.$~2000 $R_{\rm BLR}-L_{5100}$ relation is used, the probability is
0.004).  The large increase in the probability shows that the NLS1 and
BLS1 virial product distributions are more similar when one uses the
line dispersion rather than the FWHM as the H$\beta$ width measure. 
However, the two classes remain significantly different even when the
line dispersion is used and NLS1s remain with systematically smaller
virial products than BLS1s.

\subsection{Consequences on the \msig\ Plane}
\label{sec:m_sig}

The only two viable ways for NLS1s and BLS1s to both lie on the \msig\
relation are (1) for NLS1s and BLS1s to have the same black hole mass
distributions and the same stellar velocity dispersion distributions, or 
(2) for NLS1s and BLS1s to have different black hole mass distributions
(with NLS1s having lower black hole masses than BLS1s) and different
stellar velocity dispersion distributions (again, presumably with
NLS1s having smaller stellar velocity dispersions than BLS1s).  We
have shown that using the line dispersion as a measure of $\Delta V$
still produces NLS1 and BLS1 virial product distributions that are 
significantly different. For this argument, we will assume that this
virial product difference traces the distinctness of the NLS1 and BLS1
black hole mass distributions as well.  In addition, \citet{paper1}
found that NLS1s and BLS1s show no significant difference in their
distributions of stellar velocity dispersions (the K-S test probability
that the NLS1 and BLS1 stellar velocity dispersions are drawn from the
same parent population is 0.3). This conclusion is dependent on the
assumption that the width of the [\ion{O}{3}]$\,\lambda$5007 emission 
line can be used as a reliable stellar velocity dispersion indicator.
Since that question is addressed in \citet{paper3}, we assume here
that the [\ion{O}{3}] width is a fair estimator of the velocity 
dispersion in a statistical sense.  We are therefore in the situation
where the NLS1 and BLS1 black hole mass distributions are
significantly different and the stellar velocity dispersion
distributions are not significantly different.  This implies that,
even using the line dispersion as a measure of $\Delta V$, the NLS1
and BLS1 classes should lie at different locations on the \msig\
plane.

\begin{figure*}
\figurenum{3}
\plotone{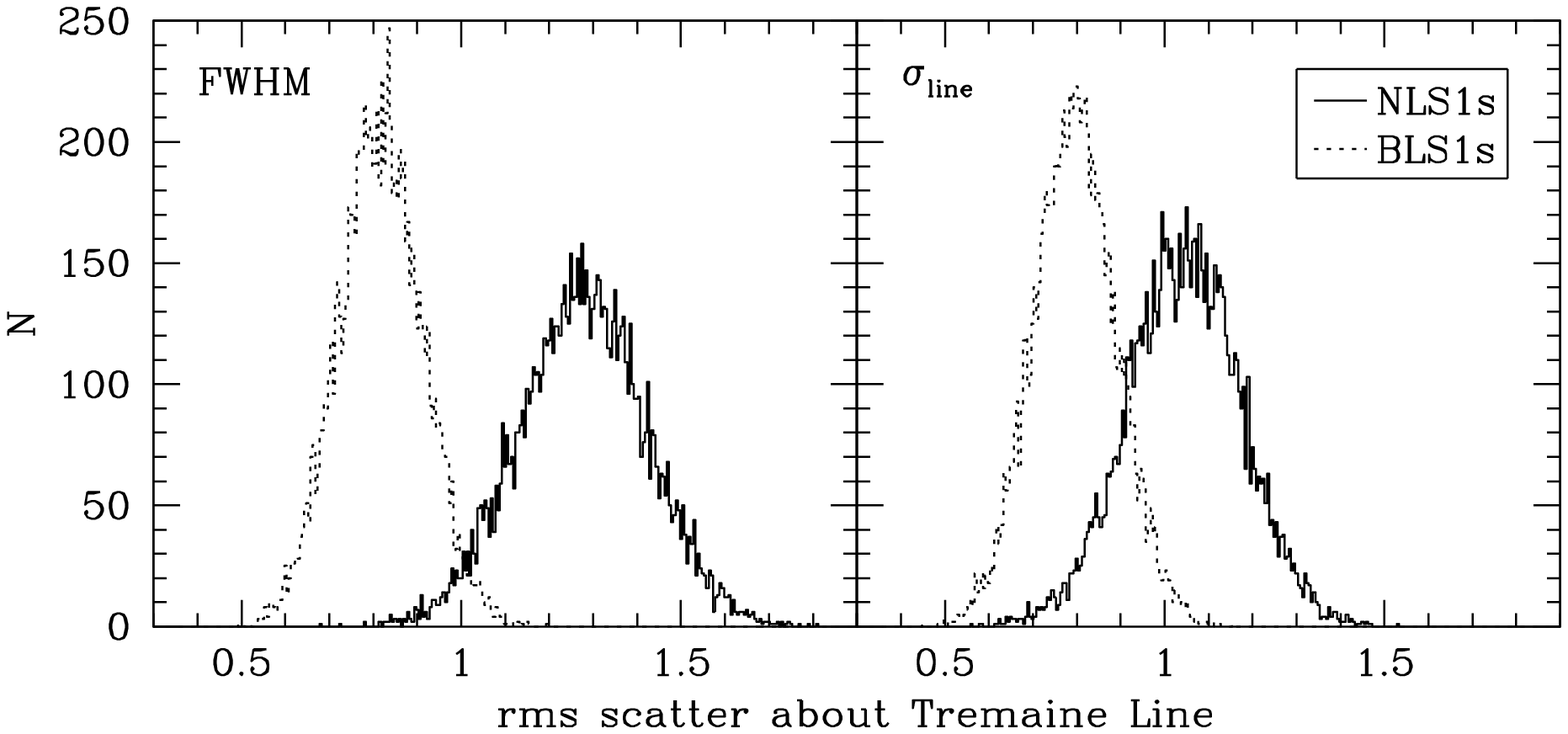}
\label{fig:rms}
\caption{Histograms showing the results of a bootstrap analysis on the
  rms scatter of our sample of NLS1s ({\it solid lines}) and BLS1s ({\it
  dotted lines}) around the \citet{tremaine} \msig\ relation.  We
  measured the rms scatter on a randomly selected (with replacement)
  sample of 40 BLS1s and 31 NLS1s from our original sample, and
  repeated the procedure 10,000 times. In the left panel, we
  calculate black hole mass using the H$\beta$ FWHM for $\Delta V$ and
  a scale factor of $f=0.53$.  In the right panel, we calculate
  black hole mass using the H$\beta$ line dispersion and a scale
  factor of $f=2.19$.  We use the width of the [\ion{O}{3}]$\,\lambda$5007
  emission line as a surrogate for the stellar velocity dispersion. \\
  }
\end{figure*}

There are many assumptions in the above argument.  We will therefore
perform two tests comparing the locations of the NLS1s and BLS1s on
the \msig\ plane.  First, we must convert our virial products into
black hole masses.  We initially used the \citet{kaspi} scale factor of
$f=0.75$ to convert our FWHM-based virial products into black hole
masses and the \citet{brad} scale factor of $f=3.85$ to convert our
line dispersion-based virial products into black hole masses.  This
procedure left a majority of the BLS1s above both the \citet{tremaine}
and the \citet{ferrarese0} fit to the \msig\ relation.  In order to
test whether the soft X-ray-selected NLS1s lie below the \msig\
relation by comparing their location to the soft X-ray-selected BLS1s,
we require our sample of BLS1s to be minimally scattered around the
\msig\ fit. Accordingly, in the remaining analysis we use masses
calculated by applying the scale factor that minimizes the rms scatter
of the BLS1s around the Tremaine line \citep[this is a modified
version of the procedure detailed in][]{onken}.  We found the best
scale factors to be $f=0.53$ to convert FWHM-based virial products
into black hole masses and $f=2.19$ to convert line dispersion-based
virial products into black hole masses.  We also completed our
analysis using the \citet{ferrarese0} fit to the \msig\ relation,
where we found $f=0.56$ to be the scale factor that best converts
FWHM-based virial products into black hole masses and $f=2.27$ to be
the scale factor that best converts line dispersion-based virial
products into black hole masses.  Because the \citet{tremaine}
relation was a better fit to our data, we use it in all relevant
figures.  We will present results based on the \citet{ferrarese0}
relation as well but note here that changing the \msig\ relation did
not significantly affect our results.

Figure~\ref{fig:msig} compares the locations of BLS1s and NLS1s
on the \msig\ plane.  The filled squares represent BLS1s, the open
squares represent NLS1s, and the solid line marks the \citet{tremaine}
fit to the \msig\ relation.  The data have been binned in log($M_{\rm
BH}/M_{\odot}$) and we have plotted the average value of log($M_{\rm
BH}/M_{\odot}$) versus the average value of log($\sigma_{\rm [O\,
III]}$) for each bin, where \sigo\ has units of kilometers per second.
The left and right panels show the locations of the AGNs on the \msig\
plane when black hole masses are calculated using the H$\beta$ FWHM
and the line dispersion, respectively. In the lower right corner of
each panel, we show typical error bars. Our calculation of the typical
error in ${\rm log} (M_{\rm BH}/M_{\odot})$ for an individual object
considers the rms scatter in the $R_{\rm BLR}-L_{5100}$ relation,
error in the measurement of the FWHM or the line dispersion, and the
unknown geometry of the BLR, which all together amounts to about 0.5
dex.  Errors for the [\ion{O}{3}]$\,\lambda$5007 emission line FWHM
are given in \citet{grupe}. Based on these values, we give a
conservative value of 0.2 dex for the error in ${\rm log} (\sigma_{\rm
[O\, III]]})$ for an individual object. This error is only the
measurement error and therefore does not include any error associated
with using the width of [\ion{O}{3}] as a surrogate for the stellar
velocity dispersion. The error bars shown in the figure are the
individual object values divided by $\sqrt{5}$, where 5 is the average
number of AGNs in each bin.

The NLS1s certainly do not appear to lie as pronouncedly below the
\citet{tremaine} line when the line dispersion is used to calculate
black hole masses.  With the large scatter in the un-binned data and the
small number of points in the binned data, we chose to use the
Mann-Whitney U test on the binned data to determine the probability that
the NLS1s and BLS1s are drawn from the same population in their $M_{\rm
BH}$ to $\sigma_{\rm [O\, III]}^{4.02}$ ratios.  Using the FWHM to
calculate black hole masses, we found the probability that the NLS1 and
BLS1 samples are drawn from the same population in their $M_{\rm BH}$ to
$\sigma_{\rm [O\, III]}^{4.02}$ ratios to be 0.001.  We also compared
the NLS1 and BLS1 $M_{\rm BH}$ to $\sigma_{\rm [O\, III]}^{4.86}$
ratios, where 4.86 is the slope of the \citet{ferrarese0} \msig\
relation; the probability remains 0.001.  Clearly, the NLS1 and BLS1
samples are different in the FWHM case. When the line dispersion is
used, the probability that the NLS1s and BLS1s are drawn from the same 
population in their $M_{\rm BH}$ to $\sigma_{\rm [O\, III]}^{4.02}$
ratios increases to 0.01 and the probability that they are drawn from
the same population in their $M_{\rm BH}$ to $\sigma_{\rm [O\,
III]}^{4.86}$ ratios increases to 0.041. The exact probability values
are sensitive to the binning parameters, but the Mann-Whitney U test
shows that there is evidence, albeit less strong  than in the FWHM
case, that NLS1s and BLS1s are drawn from different parent populations
in their $M_{\rm BH}$ to $\sigma_{\rm [O\, III]}^{4.02}$ and $M_{\rm
BH}$ to $\sigma_{\rm [O\, III]}^{4.86}$ ratios, with NLS1s having
systematically smaller values than BLS1s.

\begin{deluxetable*}{lcccccccc}
\tablewidth{0pt}
\tablecaption{The rms Scatter Values}
\tablehead{
\colhead{} &
\colhead{} &
\colhead{} &
\multicolumn{2}{c}{Tremaine} &
\colhead{} &
\colhead{} &
\multicolumn{2}{c}{Ferrarese} \\
\cline{4-5} \cline{8-9} \\
\colhead{Sample} &
\colhead{Galaxy Type} &
\colhead{} &
\colhead{FWHM} &
\colhead{$\sigma_{\rm line}$} &
\colhead{} &
\colhead{} &
\colhead{FWHM} &
\colhead{$\sigma_{\rm line}$} \\
\colhead{(1)} &
\colhead{(2)} &
\colhead{} &
\colhead{(3)} &
\colhead{(4)} &
\colhead{} &
\colhead{} &
\colhead{(5)} &
\colhead{(6)}
}
\startdata
Original ......   & NLS1s   &   & 1.29   & 1.04   &  &   & 1.41   & 1.21 \\
           & BLS1s   &   & 0.81   & 0.79   &  &   & 0.96   & 0.95 \\ \\
Bootstrap ...   & NLS1s   &   & 1.28 (0.15)   & 1.04 (0.13)   &  &    & 1.47 (0.16)   & 1.28 (0.15) \\
            & BLS1s   &   & 0.82 (0.10)   & 0.80 (0.10)  &  &    & 1.02 (0.13)   & 1.01 (0.13) \\  
\enddata
\tablecomments{Comparison of rms scatter values around the
  \citet{tremaine} and \citet{ferrarese0} fit to the \msig\ relation.
  Each rms scatter value has units of dex in log($M_{\rm
  BH}/M_{\odot}$). In the first two rows we give the rms scatter
  values of NLS1s and BLS1s when the FWHM (columns 3 and 5) or the line
  dispersion (columns 4 and 6) is used to calculate black hole masses.
  In the second two rows we provide the average rms scatter values
  from our bootstrap analysis, with standard deviation values in
  parentheses. \\ }
\end{deluxetable*}

We completed one final test comparing the locations of the NLS1s and
BLS1s on the \msig\ plane.  We simply measured the rms scatter of the
un-binned data around the \citet{tremaine} line for both the NLS1s and
the BLS1s.  Using the H$\beta$ FWHM to calculate black hole masses,
the NLS1 rms scatter is 1.29 dex and the BLS1 rms scatter is 0.81 dex
in ${\rm log}\, (M_{\rm BH}/M_{\odot})$.  Using the H$\beta$ line
dispersion to calculate black hole masses, the NLS1 rms scatter is
1.04 dex and the BLS1 rms scatter is 0.79 dex. While the BLS1 scatter
around the \citet{tremaine} line is very similar when the FWHM or the
line dispersion is used to calculate black hole masses, the NLS1
scatter is larger in the FWHM case compared to the line dispersion
case.  In other words, the NLS1s are farther from the Tremaine et
al. line in the FWHM case.  To test the significance of the rms
scatter difference between NLS1s and BLS1s in both the FWHM and line
dispersion cases, we used the bootstrap method to estimate an error in
each rms value. We used a random number generator \citep{NR} to
randomly select a sample of 31 NLS1s and 40 BLS1s from our original
sample, with replacement.  We measured the rms scatter on this new
sample and repeated the process 10,000 times. Histograms with the
results of these realizations are shown in Figure~\ref{fig:rms}, with
the solid line referring to NLS1s and the dotted line referring to
BLS1s.  The left panel of the figure shows the rms scatter histograms
when the H$\beta$ FWHM is used to calculate black hole masses.  Here
the average NLS1 rms scatter is 1.28 dex in ${\rm log}\, (M_{\rm
BH}/M_{\odot})$, with a standard deviation of 0.15 dex. The average
BLS1 rms scatter is 0.82 dex, with a standard deviation of 0.10
dex. The right panel shows the rms scatter histograms when the
H$\beta$ line dispersion is used to calculate black hole masses.  Here
the average NLS1 rms scatter is 1.04 dex in ${\rm log}\, (M_{\rm
BH}/M_{\odot})$, with a standard deviation of 0.13 dex.  The average
BLS1 rms scatter is 0.80, with a standard deviation of 0.10 dex.  The
NLS1 and BLS1 average rms scatter values differ by 3.1$\sigma$
in the FWHM case and differ by 1.9$\sigma$ in the line dispersion case
(using the NLS1 standard deviation as $\sigma$).  We also completed
the above analysis using the rms scatter around the \citet{ferrarese0}
relation and provide these results in Table~1.  Independent of the
\msig\ relation used, the results of this rms scatter test are in
agreement with the results of the Mann-Whitney U test: NLS1s and BLS1s
certainly lie in different locations on the \msig\ plane when FWHM is
used to calculate black hole masses.  In addition, NLS1s and BLS1s
remain in different locations on the \msig\ plane when the line
dispersion is used to calculate black hole masses, but the difference
is less significant.

\subsection{Eddington Ratio Comparison}
\label{sec:L_Ledd}

\begin{figure*}
\figurenum{4}
\plotone{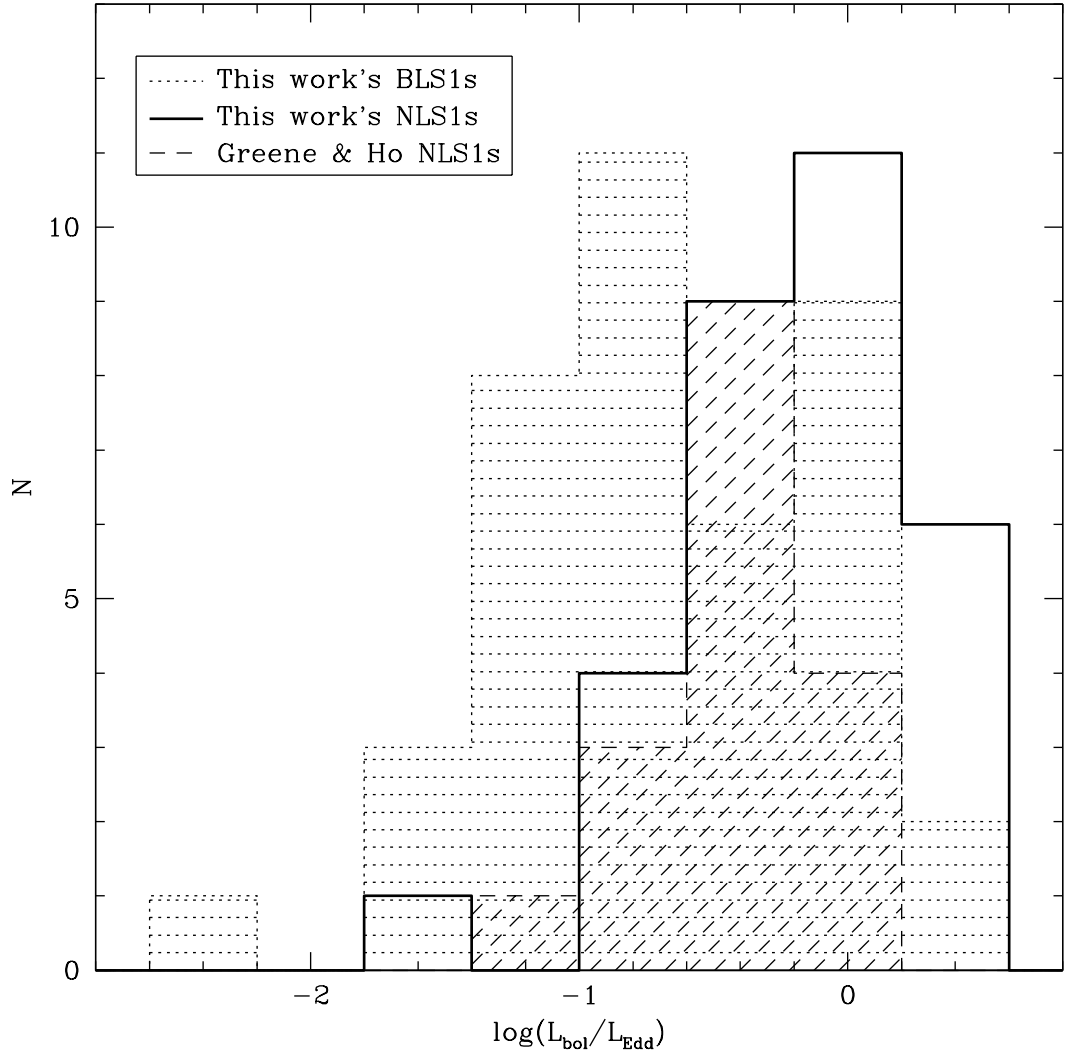}
\label{fig:L_Ledd}
\caption{Distributions of $L_{\rm bol}/L_{\rm Edd}$ for three samples:
  the soft X-ray-selected BLS1s ({\it dotted histogram}) and NLS1s
  ({\it solid line}) from our sample, with black hole mass calculated
  using the H$\beta$ line dispersion, and the optically selected
  NLS1s from Greene \& Ho (2004; {\it dashed histogram}), with black
  hole masses calculated using the H$\alpha$ FWHM. \\}
\end{figure*}

Figure~\ref{fig:L_Ledd} shows histograms of ${\rm log}(L_{\rm
bol}/L_{\rm Edd})$ for three samples: the soft X-ray-selected NLS1s of
this work ({\it solid line}), the optically selected NLS1s of Greene \& Ho
(2004; {\it dashed line}), and the soft X-ray-selected BLS1s of this work
({\it dotted line}).  One should view this figure with caution because the
black hole masses and the Eddington ratios were calculated differently
for the soft X-ray-selected AGNs and the optically selected NLS1s.  To
calculate the Eddington ratios for the soft X-ray-selected NLS1s and
BLS1s, we use the H$\beta$ line dispersion-based black hole masses.  In
contrast, \citet{greene04} use the H$\alpha$ FWHM to calculate black
hole masses for the optically selected NLS1s \citep[we have used the
corrected masses of][]{barth}.  In addition, \citet{greene04} use
$L_{\rm bol} = 9.8\lambda L_{5100}$ while \citet{grupe} used the
spectral energy distribution to estimate the bolometric luminosity of
each AGN in this work.  While the comparisons of this figure are
suspect, we present it here to show the line dispersion analog of Figure
3 of \citet{paper3}, where the H$\beta$ FWHM is used to calculate black
hole masses for the soft X-ray-selected NLS1s and BLS1s.  \citet{paper3}
found that the soft X-ray-selected NLS1s peak at the highest Eddington
ratio [mean ${\rm log}\, (L_{\rm bol}/L_{\rm Edd}) = +0.24$], the
optically selected NLS1s peak at a lower Eddington ratio [mean ${\rm
log}\, (L_{\rm bol}/L_{\rm Edd}) = -0.45$], and the soft X-ray-selected
BLS1s peak at an even lower Eddington ratio [mean ${\rm log}\, (L_{\rm
bol}/L_{\rm Edd}) = -0.75$].  When the line dispersion is used to
calculate black hole masses for the soft X-ray-selected sample, we see
that the soft X-ray-selected NLS1s have a mean ${\rm log}\, (L_{\rm
bol}/L_{\rm Edd}) = -0.19$, while the soft X-ray-selected BLS1s have a
mean ${\rm log}\, (L_{\rm bol}/L_{\rm Edd}) = -0.69$.  The trend of soft
X-ray-selected NLS1s peaking at the highest Eddington ratio, the
optically selected NLS1s peaking at a lower Eddington ratio, and the
soft X-ray-selected BLS1s peaking at an even lower Eddington ratio
remains.  However, the soft X-ray-selected NLS1s have significantly
lower Eddington ratios when the line dispersion rather than the FWHM is
used to calculate black hole masses, and thus the trend is less
pronounced.  Because we have forced the BLS1s to lie near the \msig\
relation, the BLS1s peak at similar Eddington ratios for the FWHM and
line dispersion cases.  We cannot say where the optically selected NLS1s
would lie if those black hole masses were based on the line dispersion
rather than the FWHM, but perhaps they too would be shifted towards
lower Eddington ratios.

\section{Discussion and Conclusions}
\label{sec:discussion}

To address whether the results of \citet{paper1} are affected by
substituting the H$\beta$ line dispersion for the FWHM as a measure of
the BLR gas velocity, we measured line dispersions and calculated line
dispersion-based virial products for 71 out of the 75 AGNs studied in
\citet{paper1}.  While the distributions of NLS1 and BLS1 virial
products did become significantly more similar, they remain
statistically distinct with NLS1s having smaller virial products than
BLS1s.

To examine the location of our AGNs on the \msig\ plane, we scaled
our virial products to black hole masses using the scale factor
that minimizes the rms scatter of the BLS1s around the
\citet{tremaine} fit to the \msig\ relation.  In addition, we used
\sigo\ as a surrogate for $\sigma_{\ast}$.  We found that using the line
dispersion to calculate black hole masses makes the NLS1 and BLS1
distributions significantly more similar in their locations on the
\msig\ plane.  Both our Mann-Whitney U test and our rms scatter test
show that NLS1s lie below the \msig\ relation when either the FWHM or
the line dispersion is used to calculate black hole mass.  But the
result is less significant when one uses the line dispersion.
Furthermore, the on average larger line dispersion-based black hole
masses for our sample of NLS1s leads to a lower average Eddington
ratio compared to the ratio found when FWHM is used to calculate black
hole masses.

Our results are similar to those of \citet{brad}.  Collin et al. found
that NLS1s and BLS1s require distinct scale factors to shift them onto
the quiescent galaxy \msig\ relation when virial products are
calculated using the H$\beta$ FWHM.  When the H$\beta$ line dispersion
is used, a constant scale factor is sufficient to shift both NLS1s and
BLS1s onto the \msig\ relation.  In this work, we approach the problem
by assuming a constant scale factor whether FWHM or line dispersion is
employed. \citet{paper1} found that many of their soft X-ray-selected
NLS1s lie below the \msig\ relation when virial products are
calculated using FWHM.  The authors concluded that these were highly
accreting NLS1s that had not yet achieved their ``final'' mass. In
this work, we find that using the line dispersion still leaves NLS1s
as a class below the \msig\ relation, but to a less significant degree
than when FWHM is used.

Furthermore, we agree with \citet{brad} in that FWHM is more sensitive
to some physical property of the AGN, be it perhaps the Eddington ratio
or inclination.  In addition, the NLS1s of our sample are most
affected by changing our width measure, and therefore the mystery 
physical property is likely enhanced in NLS1s.  We note here that the
\citet{brad} scale factor derived to best scale line dispersion-based
virial products into black hole masses ($f=3.85$) over-predicts the
majority of our BLS1 black hole masses with respect to the
\citet{tremaine} fit to the \msig\ relation. Even the Collin scale
factor computed using only their Population B broad-line objects
($f=3.75$) is significantly larger than our value of $f=2.19$.
However, the Collin et al. scale factor derived from FWHM measurements on
broad-line objects ($f=0.52$) is in good agreement with our value of
$f=0.53$.  The fact that our FWHM scale factor is consistent with the
Collin et al. value while our line dispersion scale factor is not could be
due to different selection effects in the optical and soft X-ray
selected samples. If this is the case, it could be giving us a clue
about the physical property that differentiates between FWHM and line
dispersion. Because of the uncertainty involved in applying the scale
factor, we trust our results comparing NLS1 and BLS1 virial products
more than the result comparing the loci of NLS1s and BLS1s on the
\msig\ plane.

We will briefly highlight differences between reverberation mapping
width measures and single-epoch width measures.  In our sample, the
fractional measurement errors of both the FWHM and the line dispersion
are about the same ($\sim0.05$).  Furthermore, the fractional
measurement error is essentially indifferent to whether an object is a
NLS1 or a BLS1. However, we could still be introducing a systematic bias
into the measurement of the FWHM or the line dispersion by using
single-epoch observations. \citet{brad04} showed that, in reverberation
mapping, one should measure the width of the emission line on the rms
spectrum, which leaves only the variable part of the spectrum. Since we
are using single-epoch observations, we must remove or avoid the
constant aspects of the spectrum.  For example, \citet{grupe} subtracted
the H$\beta$ narrow component for each AGN in our sample.  Because the
subtraction mainly affects the core of the emission line, it primarily
introduces error into our FWHM measure that would not be present in
reverberation mapping. There are also many non-variable contaminating
features surrounding H$\beta$ such as [\ion{O}{3}] and occasionally an
optically thin, very broad H$\beta$ component. These aspects of the
spectrum mainly affect the wings of the emission line and therefore
primarily introduce an error into the line dispersion that would not be
present in reverberation mapping.  These and other errors could mean
that the width measure that is best for reverberation mapping may not be
the best for single-epoch observations.  Using this data, we cannot
say which width measure is the ``right'' one.  The evidence in favor
of the line dispersion being the better choice is presented in
\citet{brad04}.

In summary, for our soft X-ray-selected sample, the virial product
distributions of NLS1s and BLS1s remain distinct when the line 
dispersion is used to measure the H$\beta$ line width; the difference,
however, is less significant than in the FWHM case.  Similarly on the
\msig\ plane, our sample of NLS1s is shifted towards the BLS1s when
the line dispersion rather than the FWHM is used; however, the NLS1s
remain below the \msig\ relation.  The disparity between the FWHM and
line dispersion results and the level of significance of the line
dispersion results are such that we cannot draw firm conclusions on
the location of soft X-ray-selected NLS1s on the \msig\ plane.  If the
scale factor to convert virial products into black hole masses is the
same for NLS1s and BLS1s, we are still left with two alternative
scenarios \citep[discussed in][]{paper3} and the present data do not
allow us choose one over the other: either (1) NLS1s lie below the
\msig\ relation indicating that their black hole masses are growing,
or (2) NLS1s lie on the \msig\ relation, so preferentially reside in
smaller mass, less luminous galaxies. In the end, more trustworthy
stellar velocity dispersions and accurate black hole mass measurements
with reverberation mapping are required for a firmer statement about
the locus of NLS1s on the \msig\ plane.  Even more basic, we must
securely determine which physical property of AGNs it is that
distinguishes FWHM from line dispersion as a BLR gas velocity
measure.

\acknowledgements
We would like to thank Bradley Peterson for helpful discussions and
for generously allowing us to use his program to calculate line 
dispersions.  We also thank the anonymous referee for useful comments.
L. W. is supported by a Graduate Fellowship from the National Science
Foundation.

\end{document}